\documentclass[rev]{jalc}
\newcommand{\shortversion}[1]{}

\newcommand{\revise}[1]{#1}

\usepackage[utf8]{inputenc}          % um Umlaute direkt eingeben zu können
\usepackage[X2,T2A]{fontenc}             % für "modernere" Schriftkodierung
\usepackage[USenglish]{babel}          % Sprachanpassungen für US-Englisch Rechtschreibung
\usepackage{lineno}
\nolinenumbers

\usepackage{xspace}

\usepackage{amsmath}
\usepackage{amsfonts}
\usepackage{mathtools}
\usepackage{amssymb}

\usepackage{comment}

\newcommand{\rt}{\mathrm{rt}}
\newcommand{\aut}{\mathcal{A}}
\newcommand{\autB}{\mathcal{B}}
\newcommand{\autC}{\mathcal{C}}
\newcommand{\Syn}{\mathrm{Syn}}
\newcommand{\MM}{\mathcal{M}}
\newcommand{\Ht}{\mathrm{ht}}

\newcommand\NP{\ensuremath{\textsf{NP}}\xspace}

\newcommand{\emptyword}{\varepsilon}

\begin{document}

\title{Shortest Synchronizing Words\\ on Product Automata}

\author[TR]{Henning Fernau\orcidID{0000-0002-4444-3220}}
\address[TR]{Universit\"at Trier, FB~4 -- Informatikwissenschaften\\ 54286 Trier, Germany.\\
\email{fernau@uni-trier.de}}

%\date{Start January 2025}

 \author[KI]{Igor Rystsov}
\address[KI]{National Technical University of Ukraine\\ Igor Sikorsky Kyiv Polytechnic Institute\\ 03056 Kiev, Ukraine.\\
\email{haryst49@gmail.com}
}

\runningauthors{\textsc{H. Fernau},  \textsc{I. Rystsov}}

\maketitle
\begin{abstract}
We consider some operations on finite automata graphs, in particular various product, union and merge operations, and investigate how they relate to the parameter~$\rt$, the \emph{length of a shortest synchronizing word}. We derive some lower and upper bounds for the operations that we study.
We also examine several classes of automata (for instance, partially ordered, commutative or idempotent automata) that are preserved under these operations and show in which way 
the parameter $\rt$ behaves in these automata classes. Also, we consider automata classes that can be constructed by combining automata from better known classes; in particular, these new classes of automata will also satisfy  the bound of the famous \v{C}ern\'y conjecture. This also motivates us to study even more closure properties. In particular, we show that certain automata classes related to synchronization form pseudo-varieties. We also commence a similar study concerning the length of shortest homing sequences of Mealy automata.
\keywords{Automata graph operations, synchronizing words, homing sequences}
\end{abstract}

\section{Introduction}

The notion of a synchronizing word (also known as a reset word) brings along one of the most intriguing open questions of (combinatorial) automata theory, namely the famous \v{C}ern\'y conjecture, see \cite{Cer64,Cer2019,Vol2008,Vol2019,VolKar2021}.\footnote{A Special Issue of the Journal of Automata, Languages and Combinatorics was dedicated to this conjecture in 2019, see \cite{Vol2019}.} In this context, the length of the shortest synchronizing word is an interesting parameter that can be associated to a deterministic finite automaton.
This parameter has also been studied from the perspective of computational complexity, e.g., showing \NP-completeness for the associated decision problem in \cite{Rys80,Epp90,GorKou95} (in chronological order), or studying approximability or parameterized tractability in \cite{Ber2014,FerBru2022,FerHegVil2015,GawStr2015,VorRom2015}.
As testified by the related so-called road coloring problem (also see \cite{Tra2009}) or also in \cite{AnaVolGus2013,Rod2019}, one can look at this problem also from the point of view of graph theory. 
In graph theory, parameters are not only studied from the viewpoint of complexity, but it is also studied how they `behave' when certain operations are applied to the graphs.
We will commence such a study for the parameter ``length of the shortest synchronizing word'', also known as ``reset threshold'', in this paper, focusing on three types of product operation on automata, a union operation and one specific merging operation. 
We will show that the maximum of the lengths of a shortest synchronizing word and its sum, respectively, of the two automata given as input to the automaton graph operation form  lower respectively upper bounds for the length of a shortest synchronizing word of the automaton obtained by this operation. While some of these results have been stated before, we are not aware of a systematic study of these graph-theoretically inspired operations.
We also look at certain subclasses of finite automata (like partially ordered or commutative) to improve on the obtained bounds for our parameter in these situations. Sometimes, the mentioned lower and upper bounds can be attained. We also show that by using these operations, we can arrive at super-classes where we know that the reset threshold is linearly bounded in the number of states. We are not aware of previous studies of such classes. Finally, we consider the question how these studies transfer to the parameter ``length of the shortest homing sequence'' that can be related to Mealy automata.

One further motivation comes from the  \v{C}ern\'y conjecture itself: if one likes to find families of automata that are even very slowly synchronizable in the sense that they might refute the conjecture, then one can restrict one's search to automata $\aut$ that cannot be further decomposed into automata $\aut_1,\aut_2$ in the sense that their product or union or merger yields $\aut$, as the linear nature of the mentioned upper bounds would mean that at least one of $\aut_1$ or $\aut_2$ would be even more slowly synchronizable. In particular when searching (possibly with the help of computers) for at least one refutation of the \v{C}ern\'y conjecture, it suffices to consider automata that are indecomposable in the sense we described.
Even if one proves the \v{C}ern\'y conjecture one day, our approach is still meaningful, as we could then try to find large\revise{, deterministically describable} classes\footnote{\revise{In a randomized setting, it is known that for binary alphabets, the length of a shortest synchronizing word of a randomly drawn DFA is growing only roughly like $\sqrt{n}$ with the number~$n$ of automata states, as shown rather recently in \cite{ChaPer2025,Mar2023}, improving on earlier works~\cite{Ber2016,Nic2019}.}} of finite automata that allow for a linear reset threshold, as explained above. In view of the practical motivations behind the notion of synchronizing words, having such a \revise{guaranteed} linear bound, i.e., really short synchronizing words, could be helpful and the question when this is possible is a natural one.

When discussing specific classes of automata, we will also prove that some of them, which are related to synchronizability, are \mbox{(pseudo-)}\linebreak[3]varieties, see \cite{BogICP99,EilSch76} giving a further link between questions on synchronizing words and algebraic automata theory.

\section{Definitions and First Results}
For a natural number $n$, let $[n]=\{1,2,\dots,n\}$ be our standard set consisting of $n$ natural numbers.

For a finite non-empty set $\Sigma$ (also called an alphabet), $\Sigma^*$ denotes the free monoid generated by $\Sigma$ with the empty word~$\emptyword$ as its neutral element; the operation is called concatenation and usually denoted by $\cdot$ or simply by juxtaposition. Elements of $\Sigma^*$ are called words (over $\Sigma$).

Throughout this paper, we consider deterministic complete finite semi-auto\-mata that can be specified as $\aut=(Q,\Sigma,\delta)$, where $Q$ is the finite set of states, $\Sigma$ is the input alphabet, and $\delta:Q\times \Sigma\to Q$ is the total transition function.
For simplicity, we refer to such an automaton as a DFA. As usual, the transition function can be extended to $\delta:Q\times \Sigma^*\to Q$ by defining $\delta(q,\emptyword)=q$ for any $q\in Q$, and by setting $\delta(q,au)=\delta(\delta(q,a),u)$ for words $u\in\Sigma^*$ of length at least one.
The function $\delta$ can be further extended to $\delta:2^Q\times \Sigma^*\to 2^Q$ by setting $\delta(S,w)=\{\delta(s,w)\mid s\in S\}$.

A DFA  $\aut=(Q,\Sigma,\delta)$ is called \emph{synchronizable} if $|\delta(Q,w)|=1$ for some word $w\in\Sigma^*$ that is known as a \emph{synchronizing word}.
We use $\Syn(\aut)$ to denote the language of synchronizing words of~$\aut$.
Let $\rt(\aut)$ denote the length of the shortest synchronizing word of the DFA~$\aut$.
For simplicity, we set $\rt(\aut)=\infty$ if and only if $\aut$ is not synchronizable. A state $q\in Q$ is called a \emph{sink state} if $\delta(q,a)=q$ for all $q\in Q$. A DFA with a sink state which is reachable from any other state is also called a \emph{0-automaton}. It is well-known that 0-automata are synchronizable.

To any DFA  $\aut=(Q,\Sigma,\delta)$, one can associate its \emph{automaton graph} $G_\aut=(Q,E)$, where the edge relation $E\subseteq Q\times Q$ is defined by $(p,q)\in E\iff \delta(p,a)=q$ for some $a\in\Sigma$. 
A state $q\in Q$ is called reachable if $(p,q)\in E^*$ for each $p\in Q$, where $E^*$ denotes the reflexive-transitive closure of the binary relation~$E$. If $a\in\Sigma$, then $\aut\downarrow a$ should denote the DFA with unary input alphabet $\{a\}$ whose transition function is obtained from $\delta$ by restriction. Hence, $G_{\aut\downarrow a}$ is a graph where all vertices have outdegree~$1$. Occasionally, we lift this notation to subalphabets $\Delta\subseteq \Sigma$, writing $\aut\downarrow\Delta$. If we denote by $G_1\sqcup G_2$ the graph obtained from $G_1=(V,E_1)$ and $G_2=(V,E_2)$ with common vertex set $V$ and edge set $E_1\cup E_2$, then $G_\aut=\bigsqcup_{a\in\Sigma} G_{\aut\downarrow a}$. As usual, we write $G_1=G_2$ if the two graphs are isomorphic.

Let $\aut_1=(Q_1,\Sigma,\delta_1)$ and  $\aut_2=(Q_2,\Sigma,\delta_2)$ be two DFA.
Then, $\aut=\aut_1\times\aut_2=(Q,\Sigma,\delta)$ describes the \emph{product automaton}, with $Q=Q_1\times Q_2$ and $\delta((q_1,q_2),a)=(\delta_1(q_1,a),\delta_2(q_2,a))$ for $a\in\Sigma$. 
This notion corresponds to the one introduced in \cite[Chapter~II]{Eil74} and is also known as \emph{direct product}.
Notice that $\aut_1\times \aut_1$ corresponds to the \emph{pair automaton} that is important in the theory of synchronizing words. Namely, $\aut_1$ is synchronizable if and only if a state of the form $(q,q)$ is reachable from any state $(p,p')$, $p\neq p'$ in $\aut_1\times \aut_1$ by using the same word.

There is an alternative way how to define the product of two automata. Now,
let $\aut_1=(Q_1,\Sigma_1,\delta_1)$ and  $\aut_2=(Q_2,\Sigma_2,\delta_2)$ be two DFA with two possibly different input alphabets and let $\Sigma=\Sigma_1\times \Sigma_2$.
Then, $\aut=\aut_1\otimes\aut_2=(Q,\Sigma,\delta)$ describes a \emph{2-tape product automaton}, reading in parallel from two input tapes, %with 
$Q=Q_1\times Q_2$ and $\delta((q_1,q_2),(a_1,a_2))=(\delta_1(q_1,a_1),\delta_2(q_2,a_2))$ for $(a_1,a_2)\in\Sigma$.

If in addition, $\Sigma_1\cap\Sigma_2=\emptyset$, we can also define $\aut=\aut_1\Box\aut_2=(Q,\Sigma,\delta)$ the \emph{sequential 2-tape product automaton} with $Q=Q_1\times Q_2$, \revise{$\Sigma=\Sigma_1\cup\Sigma_2$, and}  $\delta((q_1,q_2),a_1)=(\delta_1(q_1,a_1),q_2)$ for $a_1\in\Sigma_1$ and $\delta((q_1,q_2),a_2)=(q_1,\delta_2(q_2,a_2))$ for $a_2\in \Sigma_2$. Sometimes, we refer to~$\Box$ as the \emph{box (product) operation}. Finally, if additionally $Q_1=Q_2$, we can even consider the \emph{union automaton} $\aut=\aut_1\sqcup \aut_2=(Q,\Sigma,\delta)$ with $Q=Q_1=Q_2$ and $\delta(q,a)=\delta_i(q,a)$ if $a\in \Sigma_i$ for $i=1,2$, with $\sqcup$ being the \emph{union operation}.

There is a different interpretation of these three automata products in terms of automaton graphs. Namely, first recall two graph products, each working on two graphs $G_1=(V_1,E_1)$ and $G_2=(V_2,E_2)$:
\begin{itemize}
    \item Kronecker product $$G_1\times G_2=(V_1\times V_2,\{((x_1,x_2),(y_1,y_2))\mid (x_1,y_1)\in E_1, (x_2,y_2)\in E_2\}\,,$$
    \item Cartesian product 
    \begin{eqnarray*}G_1\Box G_2=(V_1\times V_2,\{((x_1,x_2),(y_1,y_2))&\mid\,& (x_1=y_1 \land  (x_2,y_2)\in E_2)\\& \lor\,& ( (x_1,y_1)\in E_1\land x_2=y_2) \}\,.\end{eqnarray*}
\end{itemize}
By definition, we can observe the following two interrelations when looking at the automata graphs:

\begin{proposition}\label{prop:product automaton graphs}
Let $\aut_1=(Q_1,\Sigma,\delta_1)$ and  $\aut_2=(Q_2,\Sigma,\delta_2)$ be two DFA. Then, for all $a\in\Sigma$, $G_{(\aut_1\times \aut_2)\downarrow a}=G_{(\aut_1\downarrow a)\times (\aut_2\downarrow a)}=G_{\aut_1\downarrow a}\times G_{\aut_2\downarrow a}$.
\end{proposition}

\begin{proof}Consider the situation described in the statement. Moreover, let $\aut=\aut_1\times \aut_2=(Q,\Sigma,\delta)$. Observe that the state set $Q$ of $\aut$ equals the vertex set of $G_{\aut_1\downarrow a}\times G_{\aut_2\downarrow a}$.
By definition, for all $q=(q_1,q_2)\in Q$, $\delta(q,a)=(\delta_1(q_1),\delta_2(q_2))$, i.e., $e=((q_1,q_2),(p_1,p_2))$ is an arc in $G_{(\aut_1\times \aut_2)\downarrow a}$ if and only if $e$ is an arc in $G_{(\aut_1\downarrow a)\times (\aut_2\downarrow a)}$ if and only if $e_1=(q_1,p_1)$ is an arc in $G_{\aut_1\downarrow a}$ and $e_2=(q_2,p_2)$  is an arc in $G_{\aut_2\downarrow a}$ if and only if $e$ is an arc in $G_{\aut_1\downarrow a}\times G_{\aut_2\downarrow a}$.
\end{proof}
\noindent
With a similar argument, one can prove the following two propositions.

\begin{proposition}\label{prop:2-tape product automaton graphs}
Let $\aut_1=(Q_1,\Sigma_1,\delta_1)$ and  $\aut_2=(Q_2,\Sigma_2,\delta_2)$ be two DFA. Then, for all $a_1\in\Sigma_1$ and $a_2\in\Sigma_2$, $G_{(\aut_1\otimes \aut_2)\downarrow (a_1,a_2)}=G_{(\aut_1\downarrow a_1)\otimes (\aut_2\downarrow a_2)}=G_{\aut_1\downarrow a_1}\times G_{\aut_2\downarrow a_2}$.
\end{proposition}

\begin{proposition}\label{prop:sequential 2-tape product automaton graphs}
Let $\aut_1=(Q_1,\Sigma_1,\delta_1)$ and  $\aut_2=(Q_2,\Sigma_2,\delta_2)$ be two DFA, with $\Sigma_1\cap \Sigma_2=\emptyset$. Then, for all $a_1\in\Sigma_1$ and $a_2\in\Sigma_2$,  $$G_{(\aut_1\Box \aut_2)\downarrow \{a_1,a_2\}}
= G_{(\aut_1\downarrow a_1)\Box (\aut_2\downarrow a_2)}=G_{\aut_1\downarrow a_1}\Box G_{\aut_2\downarrow a_2}\,.$$
\end{proposition}

Less common is the graph union operation of graphs with a common vertex set. But clearly, one can define $G_1\sqcup G_2=(V,E_1\cup E_2)$ for graphs $G_1=(V,E_1)$ and $G_2=(V,E_2)$. Then, the next statement is only a small observation but added for reasons of completeness.

\begin{proposition}\label{prop:union of automaton graphs}
Let $\aut_1=(Q_1,\Sigma_1,\delta_1)$ and  $\aut_2=(Q_2,\Sigma_2,\delta_2)$ be two DFA, with $\Sigma_1\cap \Sigma_2=\emptyset$ and $Q_1=Q_2$. Then,  $G_{\aut_1\sqcup \aut_2}
= G_{\aut_1}\sqcup G_{\aut_2}\,.$
\end{proposition}

When discussing special classes of automata, we will also prove that some of them connected to synchronizability are varieties. Hence, we now introduce the corresponding notions.
A class of automata  with the same input alphabet is called a \emph{variety} if it is closed under taking sub-automata, factor-automata and direct products.  A subset of states $S\subseteq Q$ is called \emph{invariant} in an automaton $\aut=(Q,\Sigma,\delta)$ if $\delta(S,a)\subseteq S$ for each $a\in \Sigma$.  In this case, we can define the \emph{sub-automaton} $\aut_1=(S,\Sigma,\delta_S)$ of $\aut$, where $\delta_S$ is a restriction of $\delta$ to the set $S\times\Sigma$. 
A partition~$P$ of states $P=\{Q_1,\dots,Q_k\}$ called a \emph{congruence} of $\aut$ if, for all $Q_i\in P$  and $a\in \Sigma$, there is a subset $Q_j\in P$ such that $\delta(Q_i,a)\subseteq Q_j$. In this case, we can define the \emph{factor-automaton} $\aut_P=(P,\Sigma,\delta_P)$ of $\aut$, where  $\delta_P(Q_i,a)=Q_j$ if  $\delta(Q_i,a)\subseteq Q_j$. \revise{It is well known and easy to prove that the class \textbf{SYN} of synchronizable automata forms a variety.}

\section{General Results}

We will first study the product operation $\times$ between automata concerning the shortest length of synchronizing words for DFA.
As this operation has been studied before in~\cite{Rom2007a}, the cited paper contains the first part of Lemma~\ref{lem:product1-synchonizability} and the second inequality of Proposition~\ref{prop:rt-product1} in its Section~5.

\begin{lemma}\label{lem:product1-synchonizability}
Let $\aut_1=(Q_1,\Sigma,\delta_1)$ and  $\aut_2=(Q_2,\Sigma,\delta_2)$ be two DFA.
$\aut_1\times\aut_2$ is synchronizable if and only if both  $\aut_1$ and $\aut_2$ are synchronizable.
More precisely, $\Syn(\aut_1\times \aut_2)=\Syn(\aut_1)\cap\Syn(\aut_2)$.
\end{lemma}

\begin{proof}
If $\aut=\aut_1\times\aut_2=(Q,\Sigma,\delta)$ is synchronizable, then there exists a word $w\in\Sigma^*$ such
that  $\delta(Q,w)$ is a singleton set, i.e., there exist $p_1\in Q_1$ and $p_2\in Q_2$ with  $\delta(Q_1\times Q_2,w)=\{(p_1,p_2)\}$. By definition of~$\delta$, this means that $\delta_1(Q_1,w)=\{p_1\}$ and $\delta_2(Q_2,w)=\{p_2\}$, i.e., $w$ is also a synchronizing word both for $\aut_1$ and for~$\aut_2$.

Conversely, if $\aut_1$ and $\aut_2$ are synchronizable, then there are words $w_1\in\Syn(\aut_1)$ and $w_2\in\Syn(\aut_2)$. Then  $w_1w_2\in \Syn(\aut_1)\cap\Syn(\aut_2)$ and so this intersection is not empty. Let  $w\in \Syn(\aut_1)\cap\Syn(\aut_2)$. Then  
by definition of~$\delta$, we see that $w$ is a synchronizing word of~$\aut$.
\end{proof}

\noindent
Looking at the proof of the preceding lemma, we can easily conclude:
\begin{lemma} Let $\aut_1=(Q_1,\Sigma,\delta_1)$ and  $\aut_2=(Q_2,\Sigma,\delta_2)$ be two DFA.
If $w_1$ is a synchronizing word for $\aut_1$ and if $w_2$ is a synchronizing word for $\aut_2$, then their concatenation $w_1w_2$ is a synchronizing word for $\aut_1\times\aut_2$.
\end{lemma}

This more explicit formulation immediately entails both upper and lower bounds on the lengths of shortest synchronizing words of $\aut_1\times\aut_2$ in relation to this parameter for $\aut_1$ and $\aut_2$:

\begin{proposition}\label{prop:rt-product1}
$\max\{\rt(\aut_1),\rt(\aut_2)\}\leq \rt(\aut_1\times\aut_2)\leq \rt(\aut_1)+\rt(\aut_2)$.
\end{proposition}

\begin{proof}
By the considerations above, it is clear that these inequalities are true if $\aut_1$ and $\aut_2$ are synchronizable. If, say, $\aut_1$ is not synchronizable, then $\rt(\aut_1)=\infty$ and hence both $\max\{\rt(\aut_1),\rt(\aut_2)\}$ and  $\rt(\aut_1)+\rt(\aut_2)$ are $\infty$. However, as $\aut_1$ is not synchronizable, also $\aut_1\times\aut_2$ is not synchronizable, so that also $\rt(\aut_1\times\aut_2)=\infty$ holds.
\end{proof} 

These bounds are best possible in general, as shown by the following two simple examples.

\begin{example}\label{exa:chains-unary}
Let $k>1$ and define $\aut_k=([k],\{a\},\delta_k)$ with $\delta_k(j,a)=\min\{j+1,k\}$ for $j\in [k]$.
For $m,n>1$, $\rt(\aut_n\times\aut_m)=\rt(\aut_{\max\{n,m\}})=\max\{n,m\}-1$. Namely, $\aut_n\times\aut_m$ and $\aut_{\max\{n,m\}}$ are isomorphic.
\end{example}

\begin{example}\label{exa:chains-binary}
Let $k>1$ and define $\aut_k=([k],\{a,b\},\delta_k)$ with $\delta_k(j,a)=\min\{j+1,k\}$ and $\delta_k(j,b)=j$ for $j\in [k]$, as well as  $\aut_k'=([k],\{a,b\},\delta_k')$ with $\delta_k'(j,b)=\min\{j+1,k\}$ and $\delta_k(j,a)=j$ for $j\in [k]$. 
For $m,n>1$, $\rt(\aut_n\times\aut_m')=\rt(\aut_n)+\rt(\aut_m')=(n-1)+(m-1)$, as we always need $n-1$ occurrences of~$a$ to synchronize the $\aut_n$-part and $m-1$ occurrences of~$b$ to synchronize the $\aut_m'$-part in any synchronizing word for $\aut_n\times\aut_m'$; in fact, any word of length $n+m-2$ that is built in this way \revise{is} a shortest synchronizing word for $\aut_n\times\aut_m'$.
\end{example}

Next, we look at the product operation $\otimes$ between automata concerning the shortest length of synchronizing words for DFA.
In the following, we will sometimes use the natural embedding morphism $\varphi$ from the free monoid $(\Sigma_1\times\Sigma_2)^*$ into the product monoid $\Sigma_1^*\times \Sigma_2^*$.

\begin{lemma} Let $\aut_1=(Q_1,\Sigma_1,\delta_1)$ and  $\aut_2=(Q_2,\Sigma_2,\delta_2)$ be two DFA.
$\aut_1\otimes\aut_2$ is synchronizable if and only if both  $\aut_1$ and $\aut_2$ are synchronizable.
\end{lemma}

\begin{proof}
If $\aut=\aut_1\otimes\aut_2=(Q,\Sigma,\delta)$ (with $\Sigma=\Sigma_1\times\Sigma_2$) is synchronizable, then there exists a word $w\in\Sigma^*$ such
that  $\delta(Q,w)$ is a singleton set, i.e., there exist $p_1\in Q_1$ and $p_2\in Q_2$ with  $\delta(Q_1\times Q_2,w)=\{(p_1,p_2)\}$. By definition of~$\delta$, this means that $\delta_1(Q_1,w_1)=\{p_1\}$ and $\delta_2(Q_2,w_2)=\{p_2\}$, where $\varphi(w)=(w_1,w_2)\in \Sigma_1^*\times \Sigma_2^*$
i.e., $w_i$ is also a synchronizing word  for $\aut_i$, with $i=1,2$.

Conversely, if $\aut_1$ and $\aut_2$ are synchronizable, then there are words $w_1\in\Sigma_1^*$ and $w_2\in \Sigma_2^*$ such that $\delta_1(Q_1,w_1)=\{q_1\}$ and $\delta_2(Q_2,w_2)=\{q_2\}$ for appropriate $q_1\in Q_1$ and $q_2\in Q_2$. Without loss of generality, assume that $\ell\coloneqq |w_1|\geq |w_2|$. Pick $a\in\Sigma_2$ arbitrarily and set $w_2'\coloneqq a^{|w_1|-|w_2|}w_2$. Then, there is a unique word $w\in\varphi^{-1}((w_1,w_2'))$.
By definition of~$\delta$, this means that $\delta(Q,w)=\delta_1(Q_1,w_1)\times \delta_2(Q_2,a^{|w_1|-|w_2|}w_2)=\{q_1\}\times\{q_2\}$, i.e., $w$ is a synchronizing word of~$\aut$.
\end{proof}

\noindent
By the construction of the previous lemma, we can conclude:
\begin{lemma}\label{lem:sw-two-tapes}
Let $\aut_1=(Q_1,\Sigma_1,\delta_1)$ and  $\aut_2=(Q_2,\Sigma_2,\delta_2)$ be two DFA.
\begin{enumerate}
    \item If $w=(w_1,w_2)$ is a synchronizing word for $\aut_1\otimes\aut_2$, then $w_1$ is  a  synchronizing word for $\aut_1$ and  $w_2$ is  a synchronizing word for~$\aut_2$.
    \item If $w_1$ is a synchronizing word for $\aut_1$ and if $w_2$ is a synchronizing word for $\aut_2$, then we find:
    \begin{itemize}
        \item if $a_1\in\Sigma_1$ and if $|w_2|\geq |w_1|$, then $(a_1^{|w_2|-|w_1|}w_1,w_2)$ or 
        \item if $a_2\in\Sigma_2$ and if $|w_2|\leq |w_1|$, then $(w_1,a_2^{|w_1|-|w_2|}w_2)$
    \end{itemize}
is a synchronizing word for $\aut_1\otimes\aut_2$.
\end{enumerate}
\end{lemma}

\begin{proposition}\label{prop:rt-product2}
$\max\{\rt(\aut_1),\rt(\aut_2)\}= \rt(\aut_1\otimes\aut_2)$.
\end{proposition}

Notice that this is the lower bound that we obtained in Proposition~\ref{prop:rt-product1}.
This could be connected from another perspective by looking at another operation, namely a restricted form of homomorphisms. The proof of the following assertion is obvious.

\begin{proposition}\label{prop:rt-hom1}
Let $\aut=(Q,\Sigma,\delta)$ and  $\aut'=(Q,\Sigma',\delta')$ be two DFA with the same state set~$Q$. 
Let $h:\Sigma\to\Sigma'$ be a mapping that we can interpret as a homomorphism mapping words over~$\Sigma$ to words over~$\Sigma'$. Further, assume that $\aut$ and $\aut'$ are compatible in the following sense: Whenever $\delta(q,a)=p$, then $\delta'(q,h(a))=p$ for any $q,p\in Q$ and $a\in\Sigma$. 
Then, we can state:
\begin{enumerate}
\item If $w$ is a synchronizing word of  $\aut$, then $h(w)$ is a synchronizing word of~$\aut'$.
    \item If $\aut$ is synchronizable, then $\aut'$ is synchronizable.
    \item $\rt(\aut)\geq \rt(\aut')$.
\end{enumerate}
\end{proposition}

Notice that one can interpret the case when $\aut$ is obtained from $\aut'$ by removing some letters (and the accordingly labeled transitions) and then taking~$h$ as the natural inclusion embedding as a special case of this proposition.

Now, let $|\Sigma|=1$ (as in Example~\ref{exa:chains-unary}). Consider two arbitrary unary DFA $\aut_1=(Q_1,\Sigma,\delta_1)$ and  $\aut_2=(Q_2,\Sigma,\delta_2)$. Now, consider $h:\Sigma\to\Sigma\times\Sigma$ with $a\mapsto (a,a)$. As $\Sigma$ is a singleton set, $h$ is bijective. Then, $\aut_1\times \aut_2$ and $\aut_1\otimes \aut_2$ are compatible in the sense of Proposition~\ref{prop:rt-hom1}. Considering the inverse $h^{-1}$ instead, one sees that also $\aut_1\otimes \aut_2$ and $\aut_1\times \aut_2$ are compatible in the sense of Proposition~\ref{prop:rt-hom1}.
This reasoning allows us to generalize Example~\ref{exa:chains-unary} as follows:

\begin{proposition}\label{prop:product-unary}
Let  $\aut_1=(Q_1,\Sigma,\delta_1)$ and  $\aut_2=(Q_2,\Sigma,\delta_2)$ be
 two unary DFA, i.e., $\Sigma=\{a\}$. Then, $$\rt(\aut_1\times\aut_2)=\rt(\aut_1\otimes\aut_2)=\max\{\rt(\aut_1),\rt(\aut_2)\}\,.$$
\end{proposition}

Let us now turn towards the sequential 2-tape automaton product and the union automaton. Now, the situation is profoundly different from that of the other two products that we examined.

\begin{proposition}
Let $\aut_1=(Q_1,\Sigma_1,\delta_1)$ and  $\aut_2=(Q_2,\Sigma_2,\delta_2)$ be two DFA with $\Sigma_1\cap \Sigma_2=\emptyset$. Setting $Q=Q_1\times Q_2$, $\Sigma=\Sigma_1\cup \Sigma_2$, let $\aut=(Q,\Sigma,\delta)=\aut_1\Box\aut_2$. Then, 
$$\rt(\aut)= \rt(\aut_1)+\rt(\aut_2)\,.$$
Moreover, $\aut_1$ and $\aut_2$ are synchronizable if and only if $\aut$ is synchronizable.
\end{proposition}

\begin{proof}
If $w_i$ is a synchronizing word for $\aut_i$ (with $i=1,2$), then $w_1w_2$ is a synchronizing word for $\aut$. Moreover, if $w$  is a synchronizing word for $\aut$, then
consider the homomorphisms $h_i$ (for $i=1,2$) with $h_i(a)=a$ if $a\in\Sigma_i$ and $h_i(a)=\emptyword$ if $a\notin\Sigma_i$. Then, $w_i=h_i(w)$ is a synchronizing word for $\aut_i$ (with $i=1,2$). This shows all claims.
\end{proof}

\begin{proposition}
Let $\aut_1=(Q_1,\Sigma_1,\delta_1)$ and  $\aut_2=(Q_2,\Sigma_2,\delta_2)$ be two DFA with $\Sigma_1\cap \Sigma_2=\emptyset$ and $Q_1=Q_2$. Setting $Q=Q_1=Q_2$, $\Sigma=\Sigma_1\cup \Sigma_2$, let $\aut=(Q,\Sigma,\delta)=\aut_1\sqcup\aut_2$. Then, 
$$\rt(\aut)\leq \min\{\rt(\aut_1),\rt(\aut_2)\}\,.$$
Hence, if $\aut_1$ and $\aut_2$ are synchronizable, then $\aut$ is synchronizable. However, the converse implication need not be true. 
\end{proposition}

\begin{proof}
If $w_i$ is a synchronizing word for $\aut_i$ (with $i=1,2$), then $w_i$ is a synchronizing word for $\aut$. This observation proves all claims but the last one. Consider $\aut_1=(\{1,2,3\},\{a\},\delta_1)$ with $\delta_1(1,a)=\delta_1(2,a)=2$ and $\delta_1(3,a)=3$, and $\aut_2=(\{1,2,3\},\{b\},\delta_2)$ with $\delta_2(1,b)=1$ and $\delta_2(2,b)=\delta_2(3,b)=2$. Obviously, neither $\aut_1$ nor $\aut_2$ is synchronizable. However, $\rt(\aut_1\sqcup\aut_2)=2$, as testified by the synchronizing word $ba$. Namely, considering the power set automaton of $\aut_1\sqcup \aut_2$, \revise{the letter}~$b$ will lead the state $\{1,2,3\}$ into $\{1,2\}$ and then $a$ will lead into $\{2\}$ as a synchronizing state.
\end{proof}

\section{Results on Special Classes of Automata}

We will discuss commutative, partially ordered, and simple idempotent automata \revise{in different subsections}. 
In connection with special classes of automata, it is also interesting to investigate if such a class is closed under the operations that we study.

In this context, the following observations are useful.

\begin{proposition}\label{prop:automata-subgraphs}
If  $\aut_1=(Q_1,\Sigma,\delta_1)$ and  $\aut_2=(Q_2,\Sigma,\delta_2)$ are DFA, then the automaton graph
$G_{\aut_1\times \aut_2}=(Q_1\times Q_2,E_\times)$ is a subgraph of $G_{\aut_1\otimes \aut_2}=(Q_1\times Q_2,E_\otimes)$. 
\end{proposition}

\begin{proof} By definition, the state sets of $\aut_1\times \aut_2$ and of $\aut_1\otimes \aut_2$ coincide, and hence the vertex sets of $G_{\aut_1\times \aut_2}$ and $G_{\aut_1\otimes \aut_2}$ are equal to $Q_1\times Q_2$.
By definition, $((q_1,q_2),(p_1,p_2))\in E_\times$ if and only if for some letter $a\in \Sigma$, $\delta_1(q_1,a)=p_1$ and $\delta_2(q_2,a)=p_2$. By definition, this means that there is a transition labeled $(a,a)$ from $(q_1,q_2)$ to $(p_1,p_2)$ in $\aut_1\otimes \aut_2$, i.e., $((q_1,q_2),(p_1,p_2))\in E_\otimes$.
\end{proof}

\subsection{Partially Ordered Automata}

A DFA $\aut=(Q,\Sigma,\delta)$ is called \emph{partially ordered} if  
%and only if 
the only possibly occurring cycles in the automaton graph $G_\aut=(Q,E)$ are self-loops. In other words, $\aut$ is  partially ordered if and only if the transitive closure $E^+$ is a partial ordering on~$Q$. Example~\ref{exa:chains-binary} gives an example of a partially ordered DFA.

\begin{theorem}\label{thm:partially-ordered-closure} Let $\aut_1=(Q_1,\Sigma_1,\delta_1)$ and  $\aut_2=(Q_2,\Sigma_2,\delta_2)$ be DFA. Then, 
$\aut_1$~and  $\aut_2$  are partially ordered if and only if $\aut_1\times \aut_2$ (with $\Sigma_1=\Sigma_2$) and  $\aut_1\otimes \aut_2$ are partially ordered.
\end{theorem}

\begin{proof}First assume that $\aut_1=(Q_1,\Sigma_1,\delta_1)$ is partially ordered.
By Proposition~\ref{prop:automata-subgraphs}, it suffices to prove that $\aut_1\otimes \aut_2$ is partially ordered, because a non-trivial cycle in $G_{\aut_1\times \aut_2}$ would be also contained in $G_{\aut_1\otimes \aut_2}$. Assume that $G_{\aut_1\otimes \aut_2}$ contains a cycle that is not a self-loop. Let $\delta$ denote the transition function of $\aut_1\otimes \aut_2$. Hence, there are states $q_1,q_2,\dots,q_i$ (with $i>1$) and symbol pairs $(a_1,b_1),(a_2,b_2),\dots,(a_i,b_i)$ such that $\delta(q_j,(a_j,b_j))=q_{k(j)}$, for $j=1,\dots,i$, with $k(j)=j+1$ if $j<i$ and $k(j)=1$ if $j=i$. By definition of~$\delta$, this means that $\delta_1(q_j,a_j)=q_{k(j)}$ and that $\delta_2(q_j,b_j)=q_{k(j)}$ for $j=1,\dots,i$. Therefore, both $G_{\aut_1}$ and $G_{\aut_2}$ have a cycle. Now observe that if both these cycles were self-loops, then also the cycle in  $G_{\aut_1\otimes \aut_2}$ that we start with would be a self-loop, contradicting our assumption.

For the converse direction,  by Proposition~\ref{prop:automata-subgraphs}, it suffices to prove that if $\aut_1\times \aut_2$ is partially ordered, then  $\aut_1=(Q_1,\Sigma_1,\delta_1)$ and  $\aut_2=(Q_2,\Sigma_2,\delta_2)$  are partially ordered, where $\Sigma=\Sigma_1=\Sigma_2$. 
We again use contraposition.
Without loss of generality, assume that $G_{\aut_1}$ contains a non-trivial cycle. This means that there is a word $u\in\Sigma^+$ of length at least two such that $\delta_1(q,u)=q$ for some $q\in Q_1$, but no proper factor $u'$ of $u$ satisfies $\delta_1(q,u')=q$. Consider now the auxiliary graph $G=(Q,E)$ obtained from $\aut_2$ by having $(r,t)\in E\iff \delta_2(r,u)=t$. As $G$ can be viewed as the automaton graph of a DFA over the one-letter alphabet $\{u\}$, we know that $G$ is a collection of directed (possible trivial) cycles with paths leading into these cycles. Observe that these cycles always correspond to non-trivial cycles in $G_{\aut_2}$. Let $s_1,\dots,s_k$ be a directed cycle in~$G$.
Let $u=a_1\cdots a_{|u|}$. Then,\\ 
\centerline{$(q,s_1)$, $(\delta_1(q,a_1),\delta_2(s_1,a_1))$, \dots, $(\delta_1(q,u),\delta_2(s_1,u))=(q,s_2)$, \dots,}
\\
\centerline{$(q,s_k)$,  $(\delta_1(q,a_1),\delta_2(s_k,a_1))$, \dots, $(\delta_1(q,u),\delta_2(s_k,u))=(q,s_1)$}
\\
is a cycle in $G_{\aut_1\times \aut_2 }$.
\end{proof}

\begin{corollary}\label{cor:PO-product-closure}
The family of partially ordered DFA is closed under the product operations $\times$ and $\otimes$.
\end{corollary}

\begin{theorem}\label{thm:partially-ordered-closure-Box} 
Let $\aut_1=(Q_1,\Sigma_1,\delta_1)$ and  $\aut_2=(Q_2,\Sigma_2,\delta_2)$ be DFA with $\Sigma_1\cap \Sigma_2=\emptyset$. Then, 
$\aut_1$~and  $\aut_2$  are partially ordered if and only if $\aut_1\Box \aut_2$ is partially ordered.
\end{theorem}

\begin{proof}
Let $\aut=\aut_1\Box\aut_2=(Q,\Sigma,\delta)$ with $Q=Q_1\times Q_2$ and $\Sigma=\Sigma_1\cup\Sigma_2$. 
If $G_{\aut_1}$ contains a non-trivial cycle, then there is a word $a_1\cdots a_n\in\Sigma_1^n$ and $n\geq 2$ pairwise different states $q_1,\dots,q_{n}$ such that $\delta_1(q_i,a_i)=q_{i+1}$ for $i\in [n-1]$ and $\delta_1(q_n,a_n)=q_1$. By definition of $\delta$, for any $p\in Q_2$, $\delta((q_i,p),a_i)=(q_{i+1},p)$ for $i\in [n-1]$ and $\delta((q_n,p),a_n)=(q_1,p)$. Hence, $G_\aut$ contains a non-trivial cycle.
Similarly, one sees that if $\aut_2$ is not partially ordered, then $\aut$ is not partially ordered. Conversely, if $\aut$ is not partially ordered, then there is a word $w=c_1\cdots c_m\in\Sigma^m$ with $m\geq 2$ that drives $\aut$ through $m$  pairwise different states. Consider the morphism $h_1:\Sigma^*\to\Sigma^*$ that maps $a$ to $a$ for $a\in\Sigma_1$ and maps letters from $\Sigma_2$ to the empty word. Similarly, $h_2:\Sigma^*\to\Sigma^*$ only preserves letters from $\Sigma_2$. Observe that $w'=h_1(w)h_2(w)$ describes a(nother) non-trivial cycle in $G_\aut$. If $|h_1(w)|,|h_2(w)|\leq 1$, then this cycle cannot be non-trivial. But if $|h_1(w)|\geq 2$, then $h_1(w)$ describes a non-trivial cycle in $\aut_1$, and similarly, if $|h_2(w)|\geq 2$, then $h_2(w)$  describes a non-trivial cycle in $\aut_2$. Hence, if $\aut$ is not partially ordered, then one of $\aut_1$ or $\aut_2$ is not partially ordered.
\end{proof}

\begin{corollary}\label{cor:PO-product-closure-Box}
The family of partially ordered DFA is closed under the operation $\Box$.
\end{corollary}

The simple examples $\aut_1=([2],\{a\},\delta_1)$ with $\delta_1(1,a)=\delta_1(2,a)=2$ and $\aut_2=([2],\{b\},\delta_2)$ with $\delta_2(1,b)=\delta_2(2,b)=1$ show that we get the following non-closure result, because $\aut_1$ and $\aut_2$ are partially ordered, while $\aut_1\sqcup \aut_2$ is not.

\begin{proposition}
The family of partially ordered DFA is not closed under the union operation~$\sqcup$.
\end{proposition}

\revise{The following statement was already proven in \cite{Ryz2019a} and in other terms in ~\cite{Rys2020}. For completeness, we give a simpler proof.}
\begin{proposition}\label{prop:PO-rt-upper-bound}
If  $\aut=(Q,\Sigma,\delta)$  is a synchronizable partially ordered DFA, then $\rt(\aut)\leq |Q|-1$.
\end{proposition}
\begin{proof}
\revise{We will prove by induction on $|Q|$. For $|Q|=1$, the proposition is obvious. Let $\aut=(Q,\Sigma,\delta)$
be a synchronizable partially ordered DFA, where $|Q|>1$, and $q$ be the maximal state in  $\aut$. Then the subset of states $Q_1=Q\backslash \{q\} $ will be invariant. By the inductive hypothesis, the subautomaton $\aut_1=(Q_1,\Sigma,\delta_1)$ possesses a synchronizing word $w$ of length $|Q_1|-1=|Q|-2$. Let  $q>\delta(q,a)$ for some $a\in \Sigma$, then $\delta(q,a)\in Q_1$. Thus the word $aw$ will be a synchronizing word for $\aut$ of length $|Q|-1$ and the proposition is proven.} 
\end{proof}
Already in the case of unary automata, for any number $k$, one can find an $n$-state automaton $\aut$ (with $n>k$) such that $\rt(\aut)=k$, because the upper bound of Proposition~\ref{prop:PO-rt-upper-bound} is attained in the unary case if the unique cycle is a self-loop, i.e., there is a path of length~$n-1$ leading to the synchronizing state, and we can find $\rt(\aut)=k<n-1$ by shortening this cycle and add further states leading to the synchronizing state in one step to reach the required number~$n$ of states.

By combining this idea with the construction of  Example~\ref{exa:chains-binary}, we obtain the following supplement to Proposition~\ref{prop:rt-product1}:

\begin{proposition}
For any numbers $k_1,k_2$ and any number $k$ with $\max\{k_1,k_2\}\leq k\leq k_1+k_ 2$, there exist partially ordered DFA $\aut_1=(Q_1,\Sigma_1,\delta_1)$ and  $\aut_2=(Q_2,\Sigma_2,\delta_2)$ with $\rt(\aut_1)=k_1$ and  $\rt(\aut_2)=k_2$ such that $\rt(\aut_1\times\aut_2)=k$.
\end{proposition}
\revise{Now we will prove a quite general statement, from which it will follow that partially ordered automata form a variety.  Recall that a class of finite monoids is called a \emph{variety} if it closed under forming finite direct products and taking submonoids and homomorphic images; see~\cite{Eil74}.}
\begin{theorem}\label{thm:monoid-variety}
\revise{If $\mathbf{V}$ is a variety of finite monoids, then the class of automata $\mathbf{A(V)}$ whose transition monoids belong to $\mathbf{V}$ is a variety of automata.}
\end{theorem}
\begin{proof}
\revise{Let $M(\aut)=\{\delta_w : w\in \Sigma^*\}$ be the transition monoid of an automaton $\aut=(Q,\Sigma,\delta)$, where $\delta_w :Q\to Q$ is the map $\delta_w(q)=\delta(q,w)$. Suppose that $\aut \in \mathbf{A(V)}$ and $M(\aut)\in \mathbf{V}$. Consider the subautomaton $\aut_1=(S,\Sigma,\delta_S)$ of $\aut$, where $S$ is an invariant subset of $Q$. Denote by $\delta_w|_S$ the restriction of a map $\delta_w$ on the set $S$ and by $M(\aut)|_S$ the restriction of the monoid of maps $M(\aut)$ on the set $S$. Then $M(\aut_1)=M(\aut)|_S$ and the mapping $\varphi: \delta_w \to \delta_w|_S$ will be a homomorphism from $M(\aut)$ to $M(\aut_1)$. Hence by definition, $M(\aut_1)\in \mathbf{V}$ and $\aut_1 \in \mathbf{A(V)}$. Thus the class $\mathbf{A(V)}$ closed under taking subautomata.}

\revise{Consider the factor-automaton $\aut_P=(P,\Sigma,\Delta)$, where $P$ is a congruence of $\aut$. Again the mapping $\varphi: \delta_w \to \Delta_w$ will be a homomorphism from $M(\aut)$ to $M(\aut_P)$. Hence by definition, $M(\aut_P)\in \mathbf{V}$ and $\aut_P \in \mathbf{A(V)}$. Therefore, the class $\mathbf{A(V)}$ is closed under taking factor-automata.}

\revise{Now consider the direct product $\aut=(Q,\Sigma,\delta)$ of automata $\aut_1=(Q_1,\Sigma,\beta)$ and $\aut_2=(Q_2,\Sigma,\gamma)$ from the class $\mathbf{A(V)}$.
By definition, the monoid $M(\aut_1)\times M(\aut_2)$ belongs to $\mathbf{V}$, since the monoids $M(\aut_1)$ and $M(\aut_2)$ belong to $\mathbf{V}$. Consider the submonoid $M=\{(\beta_w,\gamma_w):w\in\Sigma^*\}$ of the monoid $M(\aut_1)\times M(\aut_2)$ and note that $M\in \mathbf{V}$. The direct product $\beta_w \times \gamma_w$ of maps $\beta_w$ and $\gamma_w$ is defined by the formula $(\beta_w \times \gamma_w)(q_1,q_2)=((\beta_w(q_1),\gamma_w(q_2))=\delta_w(q_1,q_2)$. Hence the mapping $\varphi: (\beta_w,\gamma_w) \to \delta_w$ will be an isomorphism between monoids $M$ and $M(\aut_1\times \aut_2)$. Thus, the automaton $\aut=\aut_1\times \aut_2$ belongs to the class $\mathbf{A(V)}$.} 

\revise{By these three points, the theorem is proven.}
\end{proof}

\begin{proposition}\label{prop:ORDS-is-variety}
The class \textbf{ORDS} of partially ordered synchronizable DFA is a variety.
\end{proposition}
\begin{proof}
\revise{Brzozowski and Fich \cite{BrzFic80} proved that the transition monoid $M(\aut)$ of a partially ordered automaton $\aut$ is $R$-trivial. This means that each $R$-class of the Green equivalence $R$ in the monoid $M(\aut)$ consists of one element. Eilenberg showed \cite{Eil74a} that $R$-trivial monoids form a variety. From these results and Theorem~\ref{thm:monoid-variety}, it follows that the class \textbf{ORD} of partially ordered automata is a variety. 
%It is well known and easy to prove that the class \textbf{SYN} of synchronizing automata forms a variety. 
Then the class \textbf{ORDS} will be also a variety as an intersection of the varieties \textbf{ORD} and \textbf{SYN}.}
 \end{proof}
 
In connection with Proposition~\ref{prop:ORDS-is-variety}, it is interesting to consider the \emph{monoidal automata} that were introduced in the paper \cite{RysSzy2024}.
The full monoid of transformations on the set $Q$ we denote by $Q^Q=\{f\mid f:Q\to Q\} $. The elements of this monoid are called \emph{maps}. \revise{We denote the image of a map $f \in Q^Q$ by $im(f)=\{f(q) \mid q \in Q \}$.} The number $|im(f)|$ is called the \emph{rank} of the map $f$. A constant map is a map of rank one.  Note that a constant map transfers all states to the same state. Let $X\subseteq Q^Q$; then denote by $\langle X\rangle$ the submonoid of transformations generated by the set of maps~$X$.

A monoidal automaton (or simply automaton) is a pair $\aut=(Q,X)$, where $Q$ is a set of states and $X\subseteq Q^Q$. The monoid of maps $\langle X\rangle$ is called the transition monoid of an automaton $A$. An automaton $\aut=(Q,X)$ is synchronizable if its transition monoid $\langle X\rangle$ contains a constant map. In a 0-automaton $\aut=(Q,X)$, a constant map will be the zero element of the monoid $\langle X\rangle$. 

It is easy to see that a usual finite automaton $\aut_1=(Q,\Sigma,\delta)$ can be treated as the monoidal automaton $(Q,X)$, where $X=\{\delta_a:Q\to Q\mid a\in \Sigma \}$ and vice versa. A set of generators $X$ allows to define the measure of complexity for a map decomposition on generators, namely, let the length $l_X(f)$ of a map $f\in \langle X\rangle$ with respect to $X$ be the smallest length of a representation $f$ as a composition (product) of maps from $X$.

An automaton $\aut=(Q,X)$ is called a monoidal subautomaton of an automaton $\autB=(Q,Y)$ if $\langle X\rangle \subseteq \langle Y\rangle$. Let us consider Černý's monoidal automaton $\aut_C=(\{1,2,\cdots, n\},\{x,y\})$ where $x$ is an idempotent $x=(1,2,\cdots (n-1),1)$ \revise{(i.e., $x^2=x$, see Subsec.~\ref{subsec:idempotent})} and $y=(2,3,\cdots n,1)$ is a cyclic permutation. A map~$f$ is called nilpotent if $f^m$ is a constant map for some number $m \ge 1$. The shortest nilpotent map in Černý’s automaton is the map $y^{n-1}x$, which transfers the state $i$ into the state $i-1$, $2\le i \le n$ and leaves state $1$ in place. Then the monoidal subautomaton $\autB=(\{1,2,\cdots n\},\{w\})$ of Černý’s automaton is a synchronizable and partially ordered 0-automaton, hence from Proposition \ref{prop:PO-rt-upper-bound}, it follows that the map $w^{n-1}$ will be a constant map in~$\autB$. The constant map $(y^{n-1}x)^{n-1}$ of the length $n(n-1)$ corresponds to this composition  in Černý's automaton $\aut_C$. Therefore, to obtain the best bound we must act more carefully.

Let us consider a monoidal subautomaton $\autC=(\{1,2,\dots, n\},\{w,x\})$ of Černý’s automaton, which is also a synchronizable  partially ordered 0-automaton, since $x(n)=1$ and $x$ acts identically on the subset $Q\backslash \{n\} $.
The idempotent $x$ is also a right identity in the monoid $\langle w,x\rangle$, since $x^2=x$ and $wx=w$. So in the shortest representation of the zero map in $\autC$, the symbol $x$ can only be in the first place. We see that the composition (word) $xw^{n-2}$ of the $n-1$ will be the zero map in~$\autC$. The constant map $x(y^{n-1}x)^{n-2}$ of  length $(n-1)^2$ corresponds to this representation in~$\aut_C$.

Thus, varieties of automata can be useful for solving the Černý problem, since we can use a small upper bound for the length of constants in one automaton to obtain an upper bound for the length of constants in another automaton. This is also confirmed by the article \cite{AlmSte2009}.

\subsection{Commutative Automata}

A DFA $\aut=(Q,\Sigma,\delta)$ is called \emph{commutative} if $\delta(q,ab)=\delta(q,ba)$ for all $q\in Q$ and $a,b\in\Sigma$.
Nearly by definition, we can see that the following holds; therefore, we suppress a formal proof.

\begin{theorem}\label{thm:commutative-closure} Let $\aut_1=(Q_1,\Sigma_1,\delta_1)$ and  $\aut_2=(Q_2,\Sigma_2,\delta_2)$ be DFA. Then, 
$\aut_1$~and  $\aut_2$  are commutative if and only if $\aut_1\times \aut_2$ (with $\Sigma_1=\Sigma_2$) and  $\aut_1\otimes \aut_2$ are commutative.
\end{theorem}

\begin{theorem}\label{thm:commutative-closure-Box}Let $\aut_1=(Q_1,\Sigma_1,\delta_1)$ and  $\aut_2=(Q_2,\Sigma_2,\delta_2)$ be DFA with $\Sigma_1\cap\Sigma_2=\emptyset$. Then, 
$\aut_1$~and  $\aut_2$  are commutative if and only if $\aut_1\Box \aut_2$ is commutative.
\end{theorem}

\begin{corollary}\label{cor:commutative-closures}
The family of commutative DFA is closed under the product operations $\times$, $\otimes$, and $\Box$.
\end{corollary}

\begin{proposition}
The family of commutative DFA is not closed under the operation~$\sqcup$.
\end{proposition}

\begin{proof}
Consider the automata $\aut_1=([2],\{a\},\delta_1)$ with $\delta_1(1,a)=\delta_1(2,a)=2$ and $\aut_2=([2],\{b\},\delta_2)$ with $\delta_2(1,b)=1$ and $\delta_2(2,b)=1$. Trivially, both DFA are commutative.
Define $\aut=\aut_1\sqcup \aut_2=([2],\{a,b\},\delta)$. Then, $\delta(1,ab)=1$ while $\delta(1,ba)=2$.
\end{proof}

By induction, one can see that for commutative automata, we can assume an ordering on $\Sigma=\{a_1,a_2,\dots,a_k\}$ as $a_1<a_2<\cdots<a_k$ such that any synchronizing word first contains some occurrences of~$a_1$ (if any), then of~$a_2$, etc. We also refer to \cite{FerHof2019,Rys96,Rys97}. With this type of abbreviations, we can state an improved statement in comparison with the general one from Proposition~\ref{prop:rt-product1}. In order to formulate the next statement, let us introduce the notation $\prod_{i=1}^kw_i$ to denote $w_1w_2\cdots w_k$ in a clearer way. As we consider this in this subsection on commutative automata, even the possible misinterpretation as, e.g., $w_2w_1\cdots w_k$ does not matter.

\begin{proposition}\label{prop:rt-product1-commutative} Let $\aut_1=(Q_1,\Sigma,\delta_1)$ and  $\aut_2=(Q_2,\Sigma,\delta_2)$ be commutative DFA, with $\Sigma=\{a_1<a_2<\cdots<a_k\}$. Then, we have
$$\max\{\rt(\aut_1),\rt(\aut_2)\}\leq \rt(\aut_1\times\aut_2)\leq \sum_{i=1}^k\max\{\ell_i,\ell_i'\}\,,$$ where $\prod_{i=1}^ka_i^{\ell_i}\in\Syn(\aut_1)$ and $\prod_{i=1}^ka_i^{\ell'_i}\in\Syn(\aut_2)$ are arbitrarily chosen shortest synchronizing words.
\end{proposition}

\begin{proof}
By Proposition~\ref{prop:rt-product1}, we only have to show the (sharpened) upper bound. As mentioned above, we can assume without loss of generality that synchronizable automata digest their symbols in an arbitrary ordering.\footnote{We also refer to the discussion of jumping finite automata in \cite{FerHof2019} in this context; their behavior can be interpreted as working on the freely generated commutative monoid.}
Let $w_1=\prod_{i=1}^ka_i^{\ell_i}\in\Syn(\aut_1)$ and $w_2=\prod_{i=1}^ka_i^{\ell'_i}\in\Syn(\aut_2)$ be arbitrarily chosen shortest synchronizing words. Then, $w=\prod_{i=1}^ka_i^{\max\{\ell_i,\ell_i'\}}\in\Syn(\aut_1)\cap \Syn(\aut_2)$. By Lemma~\ref{lem:product1-synchonizability}, $w\in \Syn(\aut_1\times\aut_2)$. $|w|=\sum_{i=1}^k\max\{\ell_i,\ell_i'\}$ proves the upper bound.
\end{proof}

\begin{remark}
In the formulation of the previous proposition, we wanted to relate shortest synchronizing words in $\aut_1$ and in $\aut_2$ to synchronizing words in $\aut_1\times\aut_2$. But as we are finally aiming at finding a shortest synchronizing word in $\aut_1\times\aut_2$, it is not necessary to insist on only considering shortest synchronizing words in $\aut_1$ and in $\aut_2$. Instead, with the same arguments that have led to the previous proposition, we can prove:
$$\rt(\aut_1\times\aut_2)\leq \min\left\{\sum_{i=1}^k\max\{\ell_i,\ell_i'\}\mid \prod_{i=1}^ka_i^{\ell_i}\in\Syn(\aut_1),\prod_{i=1}^ka_i^{\ell'_i}\in\Syn(\aut_2) \right\}$$
\end{remark}

Notice that in the unary case, i.e., when $k=1$, lower and upper bound coincide as already shown in Proposition~\ref{prop:product-unary} with another argument.  

The following statement was already proven in~\cite{Rys96} and~\cite{Rys2020}. For completeness, we give a simpler proof.
\begin{proposition}\label{prop:COM-rt-upper-bound}
If  $\aut=(Q,\Sigma,\delta)$  is a synchronizable commutative DFA, then $\rt(\aut)\leq |Q|-1$.
\end{proposition}
\begin{proof}
\revise{We will prove by induction on $|Q|$. For $|Q|=1$, the proposition is obvious. Let $\aut=(Q,\Sigma,\delta)$
be a synchronizable commutative automaton, where $|Q|>1$.
There is a symbol $a\in \Sigma$ such that $\delta(Q,a)\subset Q$, since $\aut$ is synchronizable. The subset $Q_1=\delta(Q,a)$ will be invariant in $\aut$, so by the inductive hypothesis, the subautomaton $\aut_1=(Q_1,\Sigma,\delta_1)$ possesses a synchronizing word~$w$ of length $|Q_1|-1\le|Q|-2$. Thus the word $aw$ will be a synchronizing for~$\aut$ of length not greater than $|Q|-1$ and the proposition is proven.}
\end{proof}

\revise{It is evident that finite commutative monoids form a variety, therefore from Theorem~\ref{thm:monoid-variety} we have the following statement. 
\begin{proposition}\label{prop:COMS-is-variety}
The class \textbf{COMS} of commutative synchronizable DFA is a variety.
\end{proposition}
\begin{proof}
From Theorem~\ref{thm:monoid-variety}, it follows that the class \textbf{COM} of commutative automata is a variety. Then the class \textbf{COMS} will be also a variety as an intersection of the varieties \textbf{COM} and \textbf{SYN}.
\end{proof}}

\subsection{Idempotent Automata}\label{subsec:idempotent}
Let $\aut=(Q,\Sigma,\delta)$ be DFA, then an input symbol  $a\in \Sigma$ is an \emph{idempotent} if $\delta(q,aa)=\delta(q,a)$ for all $q \in Q$. This means that all states from the image $\delta(Q,a)$ are stable, i.e., $\delta(q,a)=q$  for each $q \in \delta(Q,a)$.  An idempotent symbol $a\in \Sigma$ is called \emph{simple}  if  $|\delta(Q,a)|=|Q|-1$. In this case, all states are stable except one (excluded)  state~$e$ which does not belong to  $\delta(Q,a)$. An input symbol  $a\in \Sigma$  of an automaton $\aut=(Q,\Sigma,\delta)$ is a \emph{group} (or \emph{permutation}) \emph{symbol} if  $\delta(q_1,a)\neq\delta(q_2,a)$ when $q_1 \neq q_2$. An automaton $\aut$ is called \emph{idempotent} if all its input symbols are idempotents or permutations and there is at least one an idempotent symbol, which is not a permutation. An idempotent automaton called an automaton with simple idempotents if all its idempotent symbols are simple.  

In contrast to the preceding two classes, now we have a non-closure result. It is easy to see that automata with simple idempotents are not closed under direct product. Indeed, if $\aut_1=(Q_1,\Sigma,\delta_1)$ and  $\aut_2=(Q_2,\Sigma,\delta_2)$ are automata with a simple idempotent symbol $a\in \Sigma$ and $e_1$ is  the excluded state for $a$ in~$\aut_1$, then all states $(e_1, q)$ where 
$q \in Q_2$ are not stable under the action of~$a$ in $\aut_1\times \aut_2$. So an idempotent $a\in \Sigma$ is not simple in the automaton  $\aut_1\times \aut_2$.

On the other hand, idempotent automata are closed under direct product if we assume that they share a common input alphabet $\Sigma$ which is divided into two parts $\Sigma=X \cup Y$ in the same way for both automata, a restriction for which we say that the automata \emph{share} the partition of the input alphabet. The first part $X$ contains symbols that are idempotent in both automata, and the second part $Y$ contains symbols that are permutations in both automata. Then the direct product $\aut_1\times \aut_2=(Q,X \cup Y,\delta)$ of idempotent automata   $\aut_1=(Q_1,X \cup Y,\delta_1)$ and  $\aut_2=(Q_2,X \cup Y,\delta_2)$ will be also an idempotent automaton.  We summarize these results in the following proposition.
\begin{proposition}\label{prop:idempotent automata}
The class of idempotent automata is closed under direct product of DFA with shared input alphabet partition, but the class of automata with simple idempotents is not closed under direct product.
\end{proposition}

Idempotent automata enjoy even nicer closure properties for the box product and the union operation.

\begin{proposition}\label{prop:idempotent automata-Box+union}
The class of idempotent automata is closed under the operations $\Box$ and $\sqcup$, but the class of automata with simple idempotents is closed only under the operation $\sqcup$.
\end{proposition}

\begin{proof}
In both cases, we can partition the two disjoint input alphabets $\Sigma_1$ and $\Sigma_2$ of the idempotent automata $\aut_1$ resp. $\aut_2$ into the (simple) idempotent symbols $X_1$ resp. $X_2$ and the permutation symbols $Y_1$ resp. $Y_2$. Both in $\aut_1\Box\aut_2$ and in  $\aut_1\sqcup \aut_2$, $X_1\cup X_2$ are the idempotent symbols and $Y_1\cup Y_2$ are the permutation symbols. But 
$X_1\cup X_2$ will be simple idempotent symbols only in the automaton $\aut_1\sqcup \aut_2$.
\end{proof}

However, neither the class of idempotent automata nor that of automata with simple idempotents forms a variety, as these classes are not closed under taking sub-automata.
\begin{proposition}
There exist a (simple) idempotent automaton with a sub-auto\-maton that is not idempotent.
\end{proposition}

\begin{proof}
Consider $\aut=([4],\{a,b\},\delta)$ with $\delta(1,a)=\delta(1,b)=1$, $\delta(2,a)=\delta(2,b)=2$, $\delta(3,a)=\delta(3,b)=\delta(4,b)=4$, and 
$\delta(4,a)=3$.
Then, $a$ acts as a permutation on~$[4]$, and $b$ is a simple idempotent. The sub-automaton with state set $[2]$ consists of two identity 
permutations induced by $a$ and~$b$ and is hence not an idempotent automaton.
\end{proof}
In  \cite{Rys2000}, the upper bound $2(|Q|-1)^2$ for the length of a shortest synchronizing word for an automaton with simple idempotents $\aut=(Q,\Sigma,\delta)$  was obtained. Based on the many non-closure properties of the class of DFA with simple  idempotents as exhibited above, one can easily construct larger classes of automata that also satisfy this quadratic upper bound on the reset threshold. This idea is further expanded in the next subsection.

\subsection{Combining Classes}

With our automata operations, we can also combine automata from different classes. This allows us to obtain DFA that again satisfy the \v{C}erný Conjecture. We give some examples of such combinations in this subsection.

\begin{theorem}\label{thm:po-com-combi}
There exist a partially ordered automaton $\aut_1$ and a commutative automaton $\aut_2$ such that their product (built by either $\times$ or $\otimes$) is neither partially ordered nor commutative.
\end{theorem}

\begin{proof}
It is well-known that there are DFA $\aut_1$ that are partially ordered but not commutative (as for instance $\aut_1=([3],\{a,b\},\delta_1)$ with $\delta_1(1,a)=2$, $\delta_1(1,b)=1$, $\delta_1(2,a)=\delta_1(2,b)=3$, $\delta_1(3,a)=\delta_1(3,b)=3$), and there are DFA $\aut_2$ that are commutative but not partially ordered (as for instance $\aut_2=([2],\{a,b\},\delta_2)$ with $\delta_2(1,a)=\delta_2(1,b)=2$ and $\delta_2(2,a)=\delta_2(2,b)=1$). Now, neither $\aut_1\times\aut_2$ nor $\aut_1\otimes\aut_2$ are partially ordered (by Theorem~\ref{thm:partially-ordered-closure}) or commutative (by Theorem~\ref{thm:commutative-closure}).
\end{proof}

We can do very similar constructions with the $\Box$ operation, based on Theorems~\ref{thm:partially-ordered-closure-Box} and~\ref{thm:commutative-closure-Box}. Hence, we can start considering larger classes of DFA, for instance those that can be built starting from partially ordered, commutative and idempotent automata, formed by applying any of the operations $\times$, $\otimes$, $\Box$ or $\sqcup$. By our previous considerations, we know that we get a new class of DFA that is still satisfying \revise{a quadratic (if not better) bound on the length of their synchronizing words.}

\section{Cut Vertices and Sink States}

We have seen examples of operations on automata graphs where the bounds on the length of the shortest synchronizing words related to the two original DFA can be used both to lower-bound (using the $\max$ operation) and to upper-bound (using the sum operation) the length of a shortest synchronizing word of the DFA obtained by the (product) operations. Are there any other operations on DFA that have this property? We will briefly discuss one further operation here, namely that of state identification (or merging).

\begin{definition}
Let $\aut_1=(Q_1,\Sigma,\delta_1)$ be a DFA with a sink state $s$, i.e., for all $a\in\Sigma$, $\delta_1(s,a)=s$.
Let $\aut_2=(Q_2,\Sigma,\delta_2)$ be an arbitrary DFA, with $Q_1\cap Q_2=\emptyset$. For any $q\in Q_2$, define
$\aut=(Q,\Sigma,\delta)=\aut_1\oplus_{s=q}\aut_2$, the \emph{merger} of $\aut_1$ and $\aut_2$ at $s$ and $q$,  as follows. 
$Q=(Q_1\setminus \{s\})\cup Q_2$; moreover, for any $a\in\Sigma$, define $\delta(p,a)=\delta_1(p,a)$ if $p\in Q_1$ and $\delta_1(p,a)\neq s$,  $\delta(p,a)=q$ if $p\in Q_1$ and $\delta_1(p,a)= s$, as well as $\delta(p,a)=\delta_2(p,a)$ if $p\in Q_2$.
\end{definition}

Observe that this operation is well-defined: the merger of two DFA $\aut_1$ and~$\aut_2$ is always a DFA if a sink state was selected from $\aut_1$ to perform the operation.

\begin{theorem}\label{thm:rt-merge}
Let $\aut_1=(Q_1,\Sigma,\delta_1)$ be a DFA with a sink state $s$ and let $\aut_2=(Q_2,\Sigma,\delta_2)$ be an arbitrary DFA, with $Q_1\cap Q_2=\emptyset$, and $q\in Q_2$. Then,
$$\max\{\rt(\aut_1),\rt(\aut_2)\}\leq \rt(\aut_1\oplus_{s=q}\aut_2)\leq \rt(\aut_1)+\rt(\aut_2)\,.$$
\end{theorem}

\begin{proof}
First, recall that a DFA with a sink state is synchronizable if and only if there is only one sink state in the DFA, which is reachable from any other state. 
Let us first assume that $\aut_1$ is synchronizable. Then, it is not hard to see that $\aut_1\oplus_{s=q}\aut_2$ is synchronizable if and only if $\aut_2$ is synchronizable. If $\aut_2$ is not synchronizable, then the claimed inequalities are in fact holding with equality (to $\infty$). Assume that $\aut_2$ is synchronizable. If $q$ is the synchronizing state by a shortest synchronizing word~$w$ of $\aut_2$, then $\max\{\rt(\aut_1),\rt(\aut_2)\}=\rt(\aut_1\oplus_{s=q}\aut_2)$ is only possible if $w$ also synchronizes $\aut_1$. As $q$ is a cut vertex\footnote{Recall that a vertex is called a cut vertex if its removal increases the number of connected components. In our argument, it does not matter if we reason about strongly or weakly connected components.} of the underlying automaton graph, this lower bound cannot be broken. Conversely, if $u$ synchronizes $\aut_1$ and $v$ synchronizes $\aut_2$, then $uv$ synchronizes any merger.
Finally, if $\aut_1$ is not synchronizable, then this means that there are two sink states in $\aut_1$, so that also the merger is not synchronizable, as the second sink state of $\aut_1$ would have to be the synchronizing state of the merger but it cannot be reached from any state of $\aut_2$ by construction.
\end{proof}

Notice that in contrast to the previously studied product construction, for the merger operation the upper bound given by the sum can be attained even under quite restricted conditions. This is exemplified with the following proposition.

\begin{proposition}
There is a family $\mathfrak{A}$ of partially ordered commutative DFA such that, for a well-chosen merger of two automata $\aut_1,\aut_2\in \mathfrak{A}$, we find $$\rt(\aut_1\oplus_{s=q}\aut_2)= \rt(\aut_1)+\rt(\aut_2)\,.$$
\end{proposition}

\begin{proof}
Let $\mathfrak{A}$ consist of all automata $\aut_n=([n],\{a\},\delta_n)$ with $\delta_n(i,a)=i+1$ if $i<n$ and $\delta_n(n)=n$. Obviously, $\aut_n$ is partially ordered and commutative. Moreover, $\rt(\aut_n)=n-1$. 
For $n_1,n_2\in\mathbb{N}$, consider $\aut=\aut_{n_1}\oplus_{(n_1,1)=(1,2)}\aut_2$ to be well-defined, although the state alphabets are not disjoint, as we quietly assume that the first automaton has state set $[n_1]\times[1]$ instead of $[n_1]$, while the second automaton has state set $[n_2]\times\{2\}$ instead of $[n_2]$. Then, we can even view $\aut$ as belonging to $\mathfrak{A}$, being isomorphic to $\aut_{n_1+n_2-1}$. Therefore, $\rt(\aut)=(n_1+n_2-1)-1=(n_1-1)+(n_2-1)=\rt(\aut_{n_1})+\rt(\aut_{n_2})$.
\end{proof}

\begin{remark}
Notably, we can do a similar merge construction for any state~$q$ that is a potential synchronizing state via a synchronizing word. Then, we would have to forget the outgoing edges of $q$ within~$\aut_1$ and only take the outgoing edges from~$s$ within $\aut_2$. This leads to the same bounds as stated in Theorem~\ref{thm:rt-merge}.  
\end{remark}

\section{Homing Sequences}

In this section, we will consider finite automata with outputs. For this type of automata, the notion of a homing sequence has been studied that bears some similarities with synchronizing words for finite automata without outputs; see \cite{KohJha2009,San2005}. This type of analogy can be also observed with our graph-oriented studies, as we will explain now.

A \emph{Mealy machine} $\MM=(Q,\Sigma,\Delta,\delta,\lambda)$ consists of a DFA $(Q,\Sigma,\delta)$ together with a total \emph{output function} $\lambda:Q\times \Sigma\to\Delta$.
Again, the output function is extended to work on input words by setting $$\lambda(q,a_1\cdots a_{k-1} a_k)=\lambda(q,a_1)\lambda(\delta(q,a_1),a_2)\cdots \lambda(\delta(q,a_1\cdots a_{k-1}),a_k)\,.$$
A sequence $w\in\Sigma^*$ is \emph{homing} for $\MM$ if
$$\forall p,q\in Q: (\lambda(p,w) = \lambda(q,w))\implies (\delta(p,w) = \delta(q,w))\,. $$
In particular, a synchronizing word (of the underlying DFA) is also a homing sequence, since the final state will be the same regardless of the output of an automaton.
Let us call the length of a shortest homing sequence the \emph{homing threshold} of $\MM$ and write
$\Ht(\MM)$ for short. Assuming that $\MM$ is minimized (reduced), $\Ht(\MM)\leq |Q|(|Q|-1)/2$ is known, see~\cite{Hib61}.  In contrast to synchronizing words, homing sequences always exist for reduced automata. The question if a given Mealy machine has a homing sequence of length at most~$k$ is again an \NP-hard one.

We are going to define variations of a product automaton for Mealy machines, similar to DFA. 
Let  $\MM_1=(Q_1,\Sigma,\Delta_1,\delta_1,\lambda_1)$, $\MM_2=(Q_2,\Sigma,\Delta_2,\delta_2,\lambda_2)$ be two Mealy machines. Let $Q=Q_1\times Q_2$ and $\Delta=\Delta_1\times \Delta_2$. The \emph{direct product} of $\MM_1$ and $\MM_2$ gives the Mealy machine $\MM=(Q,\Sigma,\Delta,\delta,\lambda)=\MM_1\times \MM_2$ whose underlying DFA  $(Q,\Sigma,\delta)$ is the direct product of the underlying DFA  $(Q_1,\Sigma,\delta_1)$ and  $(Q_2,\Sigma,\delta_2)$, and $\lambda((q_1,q_2),a)=(\lambda_1(q_1,a),\lambda_2(q_2,a))$. Analogously, we can also define the \emph{two-tape product} machine $\MM=(Q,\Sigma,\Delta,\delta,\lambda)=\MM_1\otimes \MM_2$, where $\MM_1=(Q_1,\Sigma_1,\Delta_1,\delta_1,\lambda_1)$, $\MM_2=(Q_2,\Sigma_2,\Delta_2,\delta_2,\lambda_2)$ are two Mealy machines. Now, $\lambda((q_1,q_2),(a_1,a_2))=(\lambda_1(q_1,a_1),\lambda_2(q_2,a_2))$. Likewise, assuming in addition $\Sigma_1\cap\Sigma_2=\emptyset$ and $\Delta_1\cap\Delta_2=\emptyset$, we can define $\MM_1\Box\MM_2$ and $\MM_1\sqcup \MM_2$, with input alphabet  $\Sigma_1\cup\Sigma_2$ and output alphabet $\Delta_1\cup\Delta_2$.

\begin{lemma} Let  $\MM_1=(Q_1,\Sigma,\Delta_1,\delta_1,\lambda_1)$ and  $\MM_2=(Q_2,\Sigma,\Delta_2,\delta_2,\lambda_2)$ be two Mealy machines.
If $w_1$ is a homing sequence for $\MM_1$ and if $w_2$ is a homing sequence for $\MM_2$, then their concatenation $w=w_1w_2$ is a homing sequence for $\MM_1\times\MM_2=(Q,\Sigma,\Delta,\delta,\lambda)$.
\end{lemma}

\begin{proof}
Let $p=(p_1,p_2),q=(q_1,q_2)\in Q=Q_1\times Q_2$ such that $ \lambda(p,w) = \lambda(q,w)$. Hence, $ \lambda_1(p_1,w) = \lambda_1(q_1,w)$ and $ \lambda_2(p_2,w) = \lambda_2(q_2,w)$.  As $w_1$ is a homing sequence for $\MM_1$, we have $ \delta_1(p_1,w_1)= \delta_1(q_1,w_1)$ and hence $ \delta_1(p_1,w)= \delta_1(q_1,w)$. As $w_2$ is a homing sequence for $\MM_2$, we have $\delta_2(\delta_2(p_2,w_1),w_2) =\delta_2(\delta_2(q_2,w_1),w_2)$, since $\lambda_2(\delta_2(p_2,w_1),w_2)=\lambda_2(\delta_2(q_2,w_1),w_2)$. Hence, $ \delta_2(p_2,w)= \delta_2(q_2,w)$ and $ \delta(p,w) = \delta(q,w)$.
\end{proof}

The lower bound can be established also easily. From the definition of the product operation, we see that a homing sequence for  $\MM_1\times\MM_2 $ will be also a homing sequence both for $\MM_1 $ and for $\MM_2 $. Therefore, we have the following result.

\begin{proposition}\label{prop:Mealy-inequalities}
Let $\MM_1$, $\MM_2$ be two Mealy machines with the same input alphabet. Then, 
$\max(\Ht(\MM_1),\Ht(\MM_2)) \leq \Ht(\MM_1\times\MM_2)\leq \Ht(\MM_1)+\Ht(\MM_2)$.
\end{proposition}

One can show that this bound can be also attained by modifying Example~\ref{exa:chains-binary}, basically by taking unary output alphabets.

For the two-tape product, an improved upper bound follows as in Lemma~\ref{lem:sw-two-tapes}.
The argument in that lemma also shows that the upper bound can be attained. Therefore, we can conclude:

\begin{proposition}
Let $\MM_1$, $\MM_2$ be two Mealy machines. Then, we have
$\Ht(\MM_1\otimes\MM_2) = \max\{\Ht(\MM_1),\Ht(\MM_2)\}$.
\end{proposition}

Similarly, one can obtain:
\begin{proposition}
Let $\MM_1$, $\MM_2$ be two Mealy machines with disjoint input alphabets and with disjoint output alphabets. Then, we have
$\Ht(\MM_1\Box\MM_2) = \Ht(\MM_1)+\Ht(\MM_2)$ and $\Ht(\MM_1\sqcup\MM_2) \leq \min\{\Ht(\MM_1),\Ht(\MM_2)\}$.
\end{proposition}
In conclusion, we find that for the product operations, reset and homing thresholds behave very similarly.
It would be interesting to find other operations on Mealy machines where one can say something about the homing threshold. Also, it could be again interesting to study sub-classes of Mealy automata.

\section{Natural Tasks and Questions}

Of course, we can discuss further automata types and more operations on automata. For instance, the operations `taking subautomata' and `quotients' have been discussed (from a different yet related viewpoint) in \cite{Vol2009}. This could be interesting in particular for operations that preserve strong connectivity of the automaton graph. Then, one could also study the \emph{avoidability threshold}, see \cite{GonJunTra2015,KisSzy2015,Vol2022}. 
But apart from this general research agenda, let us phrase some more concrete questions concerning the operations that we have studied in this paper:
\begin{itemize}
    \item Can we find general criteria for DFA, in the best case characterizations, when the two bounds of Proposition~\ref{prop:rt-product1} are attained?
    \item Can we show that each ``intermediate number'' between the lower and upper bound of Proposition~\ref{prop:rt-product1} can be attained? 
    \item Similar questions can be asked for Mealy automata (concerning homing sequences), see Proposition~\ref{prop:Mealy-inequalities}.
    \item The upper-bound construction of Proposition~\ref{prop:rt-product1} was somehow naive, as it assumed that the `work' that $w_1$ did concerning the second automaton was completely ignored. Can we always rely on first synchronizing $\aut_1$ optimally (possibly we have to pick a `most suitable' shortest synchronizing word for~$\aut_1$) and then solving a `subset synchronization' problem for $\aut_2$? 
    \item We can ask similar question for other operations that we considered.
\end{itemize}

Many automata classes are known that satisfy the upper bound on the reset threshold as conjectured by \v{C}erný and Starke, or possibly even a smaller one for a particular automata class. We can also interpret our studies on automata operators as saying that also the closure of known automata classes under these operations will then satisfy the \v{C}erný conjecture, in particular in combination. In this sense, we add further classes of automata to the already long list of automata classes satisfying  the \v{C}erný conjecture, also see (e.g.) Theorem~\ref{thm:po-com-combi} and the discussions of such a result.

More generally speaking, it would be interesting to see if graph-theoretic notions or constructions can be employed in or have bearing on other questions in automata theory.
For instance, the out-degree of an automaton graph being small testifies a combination of  ``small nondetermism'' and ``small input alphabet''. Can we obtain interesting descriptional complexity results for such graph-based measures?

\paragraph{Acknowledgements:} We would like to thank the reviewers for their valuable remarks that helped improve and extend our manuscript considerably.

\biblio{ab,jalc-sync}
\end{document}